\documentclass[twocolumn,showpacs,preprintnumbers,amsmath,amssymb]{revtex4}
\usepackage{color}
\usepackage{graphics,float}
\usepackage{graphicx}
\usepackage{dcolumn} 
\begin{document}

\title{Electronic structure and spectral properties of Am, Cm and Bk: \\
Charge density  self-consistent LDA+HIA calculations in FP-LAPW
basis}
\author{A. B. Shick${}^1$, J. Koloren\v{c}${}^{1,2}$}
\affiliation{${}^1$Institute of Physics, ASCR, Na Slovance 2, CZ-18221
Prague, Czech Republic}
\affiliation{${}^2$Department of Physics and CHiPS, North Carolina
  State University, Raleigh, North Carolina 27695, USA}
\author{A. I. Lichtenstein}
\affiliation{University of Hamburg, Jungiusstrasse 9, 20355 Hamburg,
Germany}
\author{L. Havela}
\affiliation{Department of Condensed Matter Physics, Faculty of
Mathematics and Physics, Charles University, Prague, Czech Republic}


\begin{abstract}
We provide a straightforward and numerically efficient procedure to
perform local density approximation + Hubbard I (LDA+HIA)
calculations, including self-consistency over the charge density,
within the full potential linearized augmented plane wave (FP-LAPW)
method. This implementation is all-electron, includes spin-orbit
interaction, and makes no shape approximations for the charge
density. The method is applied to calculate selected heavy actinides
in the paramagnetic phase. The electronic structure and spectral
properties of Am and Cm metals obtained are in agreement with previous
dynamical mean-field theory (LDA+DMFT) calculations and with available
experimental data. We point out that the charge density
self-consistent LDA+HIA calculations predict the $f$ charge on Bk to
exceed the atomic integer $f^8$ value by 0.22.
\end{abstract}

\date{\today}
\maketitle


\section{Introduction}

It is known that conventional band theory---local density
approximation (LDA) and its semi-local extension, generalized
gradient approximation (GGA)---%
gives poor results for actinides. Since the LDA/GGA
results are qualitatively incorrect already at the level of ground
state properties, like the equilibrium volume and magnetization, the
electronic structure theory of actinides requires that
electron-electron correlations are included beyond those given by
conventional LDA/GGA. Lately, several correlated band theory
approaches have been put forward: LDA+Hubbard U (LDA+U)
\cite{SK,shick05}, the hybrid functional (HYF) approach \cite{HYF} or
the self-interaction-corrected local spin-density (SIC-LSD)
\cite{SIC}. Each of them achieved an improvement of some particular
aspects of the electronic structure of actinides.

None of these correlated band theories has been capable to correctly
describe spectral properties of actinides. Recently, the excitations
in Pu and Am were extensively studied with the aid of a combination
of the LDA and the dynamical mean-field theory (LDA+DMFT)
\cite{gaby01,gaby,shick07,Zhu,Cris}, that successfully explains the
experimentally observed multi-peak structure in Pu valence-band
photoelectron spectra (PES). In spite of obvious progress in the
LDA+DMFT theory, it has been mostly focused on calculations of
excitations and implemented on the basis of a tight-binding
Hamiltonian built from the LDA, without self-consistency over the
charge density.

In this paper we present a simple and numerically efficient
procedure to combine the LDA+Hubbard~I approximation (HIA),
including self-consistency over the charge density, with the full
potential linearized augmented plane wave (FP-LAPW) method
\cite{Wimmer}. The FP-LAPW method makes no shape approximation for
the charge density and is considered to be state-of-the-art in
accuracy. We apply our implementation to the electronic
structure and spectroscopic properties of heavy actinides: Am, Cm
and Bk.

There is a revival of interest to the electronic and spectroscopic
properties of heavy actinides \cite{Moore}. Superconducting
temperature of Am shows complex and unconventional dependence on
lattice structure transformations \cite{Griveau}. On the basis of
standard band structure calculations it was proposed that curium is
one of a few elements that has its lattice structure stabilized by
magnetism \cite{Heathman05}. The spectroscopic studies
\cite{Moore07} suggested that 5$f$ states of Cm are shifted towards
the LS coupling limit, unlike most actinide elements where the
intermediate coupling prevails.

The paper is organized as follows. For the sake of completeness, in
Sec.~II we recall the basic equations of the LDA+DMFT in a formulation
of Ref.~\cite{LK99}. Then we describe charge density self-consistent
LDA+HIA approximation implemented in FP-LAPW method. In Sec. III we
present the results of the charge density self-consistent LDA+HIA
calculations for $5f$ Am, Cm, and Bk elemental metals in the
paramagnetic state. These results are compared with previous work
and additional features are pointed out.

\section{Methodology}
We start with the multi-band Hubbard Hamiltonian \cite{LK99}
$H = H^0 + H^{\rm int} $, where
\begin{eqnarray}
\label{eq:1ph} H^0 &=& \sum_{i,j} \sum_{\gamma_1, \gamma_2} H^0_{i
\gamma_1, j \gamma_2}
                 c^{\dagger}_{i \gamma_1} c_{j \gamma_2} \nonumber \\
    &=& \sum_{\bf k} \sum_{\gamma_1, \gamma_2} H^0_{\gamma_1, \gamma_2} ({\bf k})
        c^{\dagger}_{\gamma_1}({\bf k}) c_{\gamma_2}({\bf k})
\end{eqnarray}
is the one-particle Hamiltonian found from ab initio electronic
structure calculations of a periodic crystal, including the
spin-orbit coupling (SOC). The indices $i,j$ label lattice sites,
$\gamma = (l m \sigma)$ denote spinorbitals $\{ \phi_{\gamma}
\}$, and ${\bf k}$ is a k-vector from the first Brillouin zone. It
is assumed that the electron-electron correlations between $s$, $p$, and
$d$ electrons are well described within the density functional theory,
while the correlations between the $f$ electrons have to be considered
separately by introducing the interaction Hamiltonian
\begin{multline}
\label{eq:hint} H^{\rm int} = \frac{1}{2} \sum_{i}
\sum_{m_1,m_2,m_3,m_4}^{\sigma,\sigma'}
  \langle m_1,m_2|V_{i}^{ee}|m_3,m_4 \rangle \\
  \times c_{i m_1 \sigma}^{\dagger} c_{i m_2 \sigma'}^{\dagger}
  c_{i m_4 \sigma'} c_{i m_3 \sigma}  \, .
\end{multline}
The operator $V^{ee}$ represents an effective on-site Coulomb
interaction \cite{LK99} expressed in terms of the Slater integrals
$F_k$ and the spherical harmonics ${|lm \rangle}$.

In what follows we use a local approximation for the one-particle
selfenergy $\Sigma({\bf k},z)$ which contains the electron-electron
correlations, i.~e., we assume that the selfenergy is site-diagonal and
therefore independent of ${\bf k}$. The corresponding one-particle
Green function reads
\begin{eqnarray}
\label{eq:1gf} G({\bf k},z) = \Big( z + \mu - H^0 ({\bf k}) -
\Sigma(z) \Big)^{-1} \, ,
\end{eqnarray}
where $z$ is a (complex) energy measured with respect to the chemical
potential $\mu$. The interaction term, Eq.~(\ref{eq:hint}), acts only in the
subspace of $f$-states. Consequently, the selfenergy $\Sigma(z)$ is
nonzero only in the subspace of the $f$-states.

The self-consistent procedure to solve the periodic lattice problem
in the DMFT approximation is now formulated in the usual way making
use of the ``impurity'' method of Ref.~\cite{LK99}. The DMFT
self-consistency condition is achieved by equating the local Green
function in a solid to the Green function of a single-impurity
Anderson model (SIAM) that describes an isolated multiorbital
impurity surrounded by a bath of uncorrelated delocalized electrons.
Starting with the single-particle Hamiltonian $H^0({\bf k})$ and a
guess for the local selfenergy $\Sigma(z)$, the local Green function is
calculated by integrating $G({\bf k},z)$, Eq.~(\ref{eq:1gf}), over
the Brillouin zone. Subsequently, bath Green function (the so-called
Weiss field) is calculated \cite{LK99}, and used to solve the SIAM.
New local $\Sigma(z)$ is evaluated, which is inserted back into
Eq.~(\ref{eq:1gf}). In addition, the charge density needed to
construct the single-particle Hamiltonian $H_0({\bf k})$ in
Eq.~(\ref{eq:1gf}) has to be calculated self-consistently from the
local Green function.

\subsection{Hubbard-I approximation for $\Sigma(z)$}
We make use of the multiorbital HIA, which is suitable for
incorporating the multiplet transitions into the electronic
structure, as it is explicitly based on the exact diagonalization of
an isolated atomic-like shell. Further, we restrict our formulation
to the paramagnetic phase. In HIA, only the site-diagonal terms from
one-particle Hamiltonian Eq.~(\ref{eq:1ph}) are
retained~\cite{LK99}, and the on-site atomic-like Hamiltonian
including SOC is constructed, see also Ref.~\cite{shick07},
\begin{align}
H^{\rm at}  = & \sum_{m_1,m_2}^{\sigma, \sigma'} \xi ({\bf l} \cdot
{\bf s})_{m_1 m_2}^{\sigma \; \; \sigma'}
 c_{m_1 \sigma}^{\dagger}c_{m_2 \sigma'} \nonumber \\
&+ \frac{1}{2} \sum_{m_{1}...m_{4}}^{\sigma, \sigma'}
\langle m_1 m_2|V^{ee}|m_3 m_4 \rangle \nonumber \\
& \qquad\qquad\qquad\times c_{m_1 \sigma}^{\dagger}
c_{m_2\sigma'}^{\dagger} c_{m_4 \sigma'} c_{m_3 \sigma} \; , \label{eq:10}
\end{align}
where $\xi$ is the SOC parameter. Note that  the crystal field terms
are not included in Eq.~(\ref{eq:10}), and will be treated on the LDA
level that is sufficient for our applications.

Consecutively, exact diagonalization, $H^{\rm at} |\nu \rangle =
E_\nu |\nu\rangle $, is performed in order to obtain all eigenvalues
$E_\nu$ and eigenvectors $|\nu \rangle$ that are used to calculate
the atomic Green function
\begin{multline}
[G^{\rm at}(z)]_{\gamma_1 \gamma_2} = \frac{1}{Z} \,
\sum_{\nu,\mu} \frac{\langle \mu|c_{\gamma_1}|\nu \rangle \langle
\nu|c_{\gamma_2}^{\dagger}|\mu \rangle} {z +
E_\mu-E_\nu+\mu_H} \\
\times \Bigl[e^{-\beta (E_\nu - \mu_H N_\nu)}
         + e^{-\beta (E_\mu - \mu_H N_\mu)}\Bigr] \, .
\label{eq:12}
\end{multline}
Here $\beta$ is the inverse temperature, $Z$ is the partition
function, and $N_{\nu}$ is the number of particles in the state
$|\nu\rangle$. These $N_{\nu}$ are eigenvalues of the particle
number operator that commutes with the atomic Hamiltonian
Eq.~(\ref{eq:10}). Parameter $\mu_H$ plays a role of the HIA chemical
potential. Actual choice of $\mu_H$ will be discussed later.
Finally, the atomic self-energy is evaluated as
\begin{multline}
\Big[\Sigma_{H}(z)\Big]_{\gamma_1 \gamma_2} =
z\delta_{\gamma_1 \gamma_2}\\
- \biggl[\, \xi ({\bf l} \cdot {\bf s})+
\Big(G^{\rm at}(z)\Big)^{-1}\biggr]_{\gamma_1 \gamma_2} .
\label{eq:13}
\end{multline}
This $\Sigma_{H}(z)$
contains all local spin-orbit and Coulomb correlation effects.

\subsection{Self-consistency over charge density:
\\
Local Density Matrix Approximation}

Instead of solving Eq.~(\ref{eq:1gf}) directly, we look for an
approximate solution including charge density self-consistency in a
way which is similar to the well known rotationally invariant LDA+U
method \cite{LAZ95}.

We start with calculating the HIA $\Sigma(z)$, Eq.~(\ref{eq:13}), for
given $\mu_{H}$. In our applications, this starting $\mu_{H}$
corresponds to the nominal atomic $f$-shell occupation $n_f$. The
initial solution for lattice electrons is represented by the LDA Green
function matrix in the local basis $\{ \phi_{\gamma} \}$,
\begin{multline}
\Big[G_{LDA}(z)\Big]_{\gamma_1 \gamma_2} =  \\[.2em]
\frac{1}{V_{\rm BZ}} \int_{\rm BZ}{\rm d}{\bf k}
\Big[z+\mu-H_{LDA}({\bf k}) \Big]^{-1}_{\gamma_1 \gamma_2} \, .
\label{eq:14}
\end{multline}
Note that the SOC is included in the LDA Hamiltonian
$H_{LDA}({\bf{k}})$. The local impurity Green function
is calculated combining $\Sigma(z)$ and
$G_{LDA}(z)$,
\begin{multline}
\Big[G(z)\Big]^{-1}_{\gamma_1 \gamma_2}  =
\Big[{G}_{LDA}(z)\Big]^{-1}_{\gamma_1 \gamma_2}  \\
 -  \biggl(\Delta \epsilon \delta_{\gamma_1 \gamma_2} +
\Big[\Sigma_{H}(z)\Big]_{\gamma_1 \gamma_2}
\biggr) \; ,
\label{eq:16}
\end{multline}
where $\Delta \epsilon$ is chosen to keep the given number of
$f$-electrons $n_f$, and serves as an analogon of the difference
between the impurity and the lattice chemical potentials
\cite{shift}.

With the aid of $G(z)$ from Eq.~(\ref{eq:16}),  the occupation
matrix {$n_{\gamma_1 \gamma_2} = -\pi^{-1}\mathop{\rm Im} \int^{E_F}
{\rm d} z \, [G(z)]_{\gamma_1 \gamma_2}$} is evaluated , and used to
construct the effective ``LDA+U potential'' \cite{shick2004},
${V}_{U} = \sum_{\gamma_1 \gamma_2}
  |\phi_{\gamma_1} \rangle V_{U}^{\gamma_1 \gamma_2} \langle
  \phi_{\gamma_2}|$,
where
\begin{align}
 V_{U}^{\gamma_1 \gamma_2} =& \sum_{\gamma \gamma'}
 \Big(\langle \gamma_2 \gamma  |V^{ee}|\gamma_1  \gamma' \rangle
 - \langle \gamma_2 \gamma |V^{ee}| \gamma'  \gamma_1 \rangle
 \Big)
n_{\gamma \gamma'} \nonumber\\
&- V_{dc} \delta_{\gamma_1 \gamma_2}\,.
\label{eq:ldau}
\end{align}
In what follows, we have adopted the fully-localized (or
atomic-like) limit (FLL) prescription of Solovyev {\em et
  al.}~\cite{Solovyev} for the double-counting term $V_{dc} =
U (n_f-1/2) -
J(n_f-1)/2$.

The set of  Kohn-Sham-like equations is solved self-consistently
over the charge density  $\rho(\mathbf{r})$
\begin{gather}
\Big( -\nabla^{2} + V_{LDA}(\mathbf{r}) + V_{U} + \xi ({\bf l}
\cdot {\bf s})  \Big)  \Phi_i({\bf r}) = e_{i}
\Phi_{i}({\bf r}) \;, \nonumber \\
\rho(\mathbf{r}) = \sum_i^{occ} \Phi^{\dagger}_i({\bf
r})\Phi_i({\bf r}) \, ,
\label{eq:kseq}
\end{gather}
where the effective potential is the sum of the standard LDA
potential $V_{LDA}(\mathbf{r})$ and the on-site electron-electron
interaction potential $V_{U}$.
Solving Eqs.~(\ref{eq:kseq}) is similar to solving Eq.~(\ref{eq:1gf})
in a sense that the selfenergy matrix $\Sigma(z)$ from
Eq.~(\ref{eq:13}) is substituted by the energy independent potential
matrix defined in Eq.~(\ref{eq:ldau}). After the self-consistency
over the charge density is achieved, the LDA+U Green function matrix
$G_{U}(z)$ in the local basis $\{ \phi_{\gamma} \}$ is calculated
from Eq.~(\ref{eq:14}), substituting $H_{LDA}$ by LDA+U Hamiltonian.
Finally, new uncorrelated Green function
\begin{eqnarray}
{G}_{LDA}(z) = \Big[ G_{U}^{-1}(z) + V_{U}(z) \Big]^{-1}
\label{eq:15}
\end{eqnarray}
is evaluated. The self-consistency loop is closed by inserting this
new
${G}_{LDA}(z)$ into the matrix equation~(\ref{eq:16}). In addition,
an updated selfenergy $\Sigma(z)$ is calculated with the aid of
Eqs.~(\ref{eq:12},\ref{eq:13}), where the new value of $\mu_H$ is
set equal to the double-counting potential $V_{dc}$ that corresponds
to $n_f$ obtained from the LDA+U Green function.

The condition $\mu_H=V_{dc}$ is essential and can be justified as
follows. The double-counting term $V_{dc}$ accounts approximately
for the electron-electron interaction energy $E^{ee}_{LDA}$ already
included in the LDA. Namely, $V_{dc}$ is a derivative of this mean
energy contribution with respect to the $f$-shell occupation $n_f$,
$V_{dc}=\partial E^{ee}_{LDA}/\partial n_{f}$. Indeed, it represents
a mean-field value of the chemical potential $\mu_H$ that controls
the number of $f$ electrons.

The FLL \cite{Solovyev, LAZ95} choice of the double counting
$V_{dc}$ is not unique and other prescriptions, for instance the
so-called ``around-mean-field'' $V_{dc}$ \cite{Anisimov,Kunes}, can
be used. Up to date, there is no precise solution for the double
counting in the conventional LDA/GGA as it does not have a
diagrammatic representation that would provide explicit
identification of the corresponding many-body interaction terms.
Therefore, ``physical'' arguments prevail in the choice of $V_{dc}$.
Since we will be dealing with heavy actinides with well localized
$f$-manifolds, it is reasonable to use the FLL double counting that
is assumed to perform better for the case of $f$-occupation close to
integer.

We will refer to our procedure as the ``local density matrix
approximation'' (LDMA), since full convergence for ${G}_{LDA}$,
$\rho(\mathbf{r})$ and $\mu_H$ is achieved when the local occupation
matrix $n_{\gamma_1 \gamma_2}$ is converged. We would like to emphasize that
the self-consistency condition of equating the occupation matrix
obtained from the local impurity Green function Eq.~(\ref{eq:16}) to
the local occupation matrix in solid (used in the LDA+U potential
Eq.~(\ref{eq:kseq})) is
a subset of general DMFT condition that the
SIAM Green function is equal to the local Green function in a solid
\cite{Georges}.

What makes our approach different from the conventional LDA+HIA
given by Eq.~(\ref{eq:1gf}) and from similar basis set extension method
of Ref.~\cite{Savrasov06}, is that we interchange the ``inner'' DMFT
self-consistency loop over the bath Green function ${G}_{LDA}$,
Eq.~(\ref{eq:15}), and the ``outer'' self-consistency loop over the
charge density $\rho(\mathbf{r})$, Eq.~(\ref{eq:kseq}).

Up to now, our considerations did not depend on the choice of the
basis set. The method becomes basis dependent, when a projector for
the Bloch state $\Phi_i({\bf r})$ solution of Eq.~(\ref{eq:kseq}) on
the local basis $\{ \phi_{\gamma} \}$ is specified. The FP-LAPW
method uses a basis set of plane waves that are matched onto a
linear combination of all radial solutions (and their energy
derivative) inside a sphere centered on each atom. In this case, we
make use of the projector technique which is described in detail in
Ref.~\cite{SLP99}. It is important to mention that due to the full
potential character care should be taken \cite{shick2004} to exclude
the double-counting of the $f$-states non-spherical contributions to
the LDA and LDA+U parts of potential in Eq.~(\ref{eq:kseq}).

\section{Results}
As representative systems to illustrate the LDMA numerical
procedure, we select heavy actinides---Am, Cm and Bk. For all of them,
the HIA is expected to provide a reasonable approximation for the
selfenergy. We focus on comparison between the theory and available
experimental results for valence-band photoelectron spectra (PES) as
well as X-ray absorption (XAS) and electron energy-loss (EELS)
spectroscopies. This comparison is often taken as important
criterion of truthfulness of electronic structure calculations.

Experimental valence-band PES spectra will be compared with valence
spectral densities resulting from the self-consistent LDMA.  For the XAS
and EELS experiments, we will compare the branching ratio $B$ as
well as strength of the spin-orbit coupling $w^{110}$ for
core-to-valence $4d$--$5f$ transition,
\begin{eqnarray}
w^{110} &=& n_f^{7/2} -\frac{4}{3} n_f^{5/2}\,,  \nonumber  \\
\frac{w^{110}}{(14 - n_f)} - \Delta &=& -\frac{5}{2} \left( B -
\frac{3}{5} \right) \, , \label{eq:BR}
\end{eqnarray}
where $\Delta$ represents a small correction term \cite{Moore}.

The actinides were calculated assuming paramagnetic state with
$fcc$-crystal structure and the experimental volume per atom.
The parameters of the local Hamiltonian, Eq.~\eqref{eq:10}, were
chosen as follows: $U=F_0=4.5$~eV, $F_2$, $F_4$ and $F_6$ were taken
from Table III of Ref.~\cite{Moore}, and values of the SOC parameter
$\xi$ were extracted from LDA calculations ($\xi_{Am}=0.35$~eV,
$\xi_{Cm}=0.36$~eV and $\xi_{Bk}=0.42$~eV). The HIA Green function
and selfenergy, Eqs.~(\ref{eq:12}) and~(\ref{eq:13}), were
calculated along the real axis $z = \mathop{\rm Re}z + i\delta$ with
$\delta=0.1$~eV. In the process, values $10$~eV$^{-1}$ and
$100$~eV$^{-1}$ were used for the inverse temperature $\beta$. For
self-consistency, 108 special k-points \cite{BZ} in the irreducible
1/8th part of the BZ were used. The same sphere radius $R_{MT} =
3.1$ a.u. was used for all actinides, and $R_{MT} \times K_{max} =
10.70$ determined the basis set size. The $f$-manifold occupation
$n_f$ is varied in the calculations until the convergence better
than $0.01$ for $n_f$ and $0.001$ for all components of the on-site
occupation matrix $n_{\gamma_1 \gamma_2}$ is achieved. The charge
density is fully converged to better than $10^{-5}$ e/a.u.$^3$ at
each iteration.

\begin{figure}[h]
\vspace*{-0.5in}
\includegraphics[width=3.25in,clip]{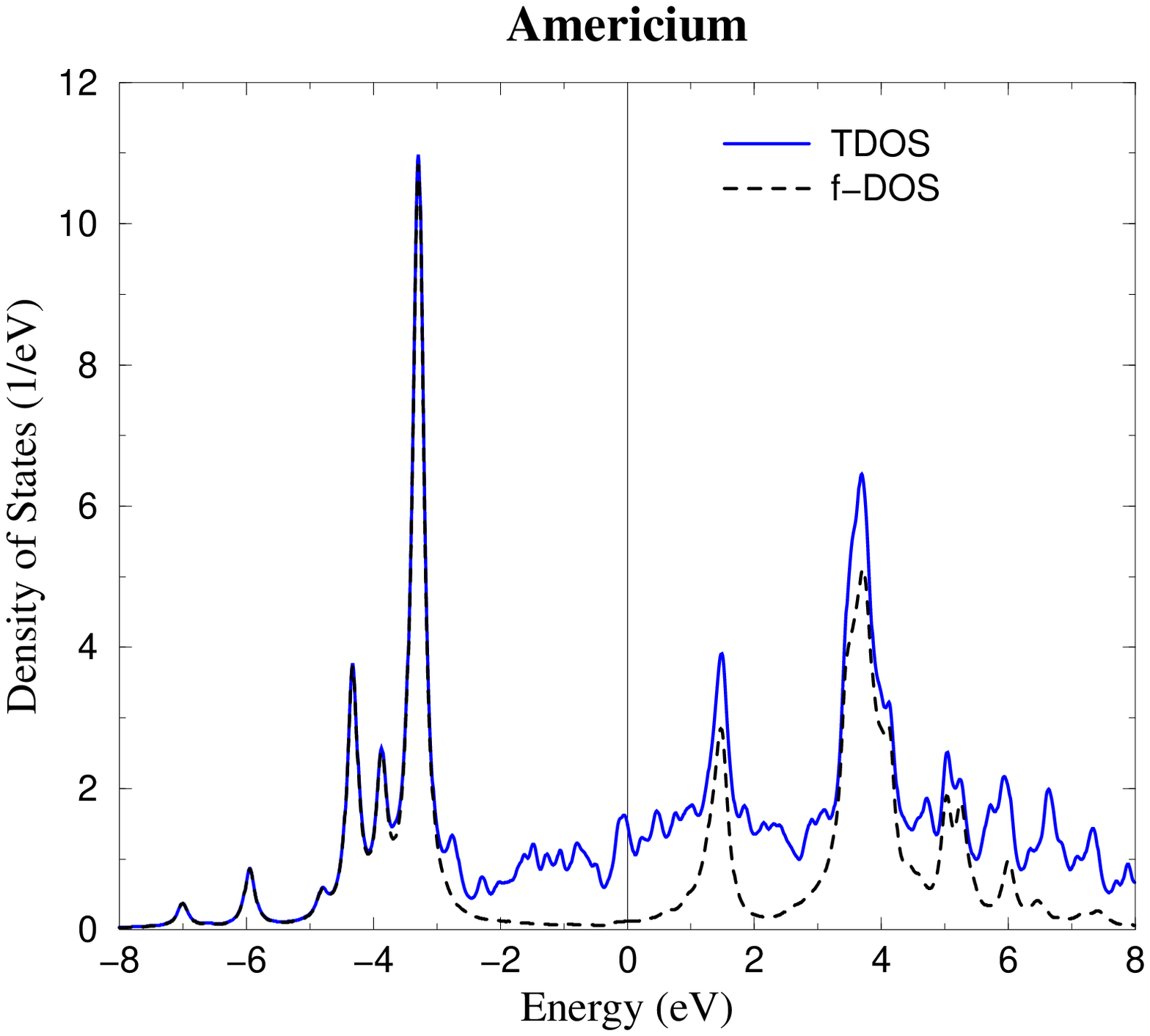}
\includegraphics[width=3.25in,clip]{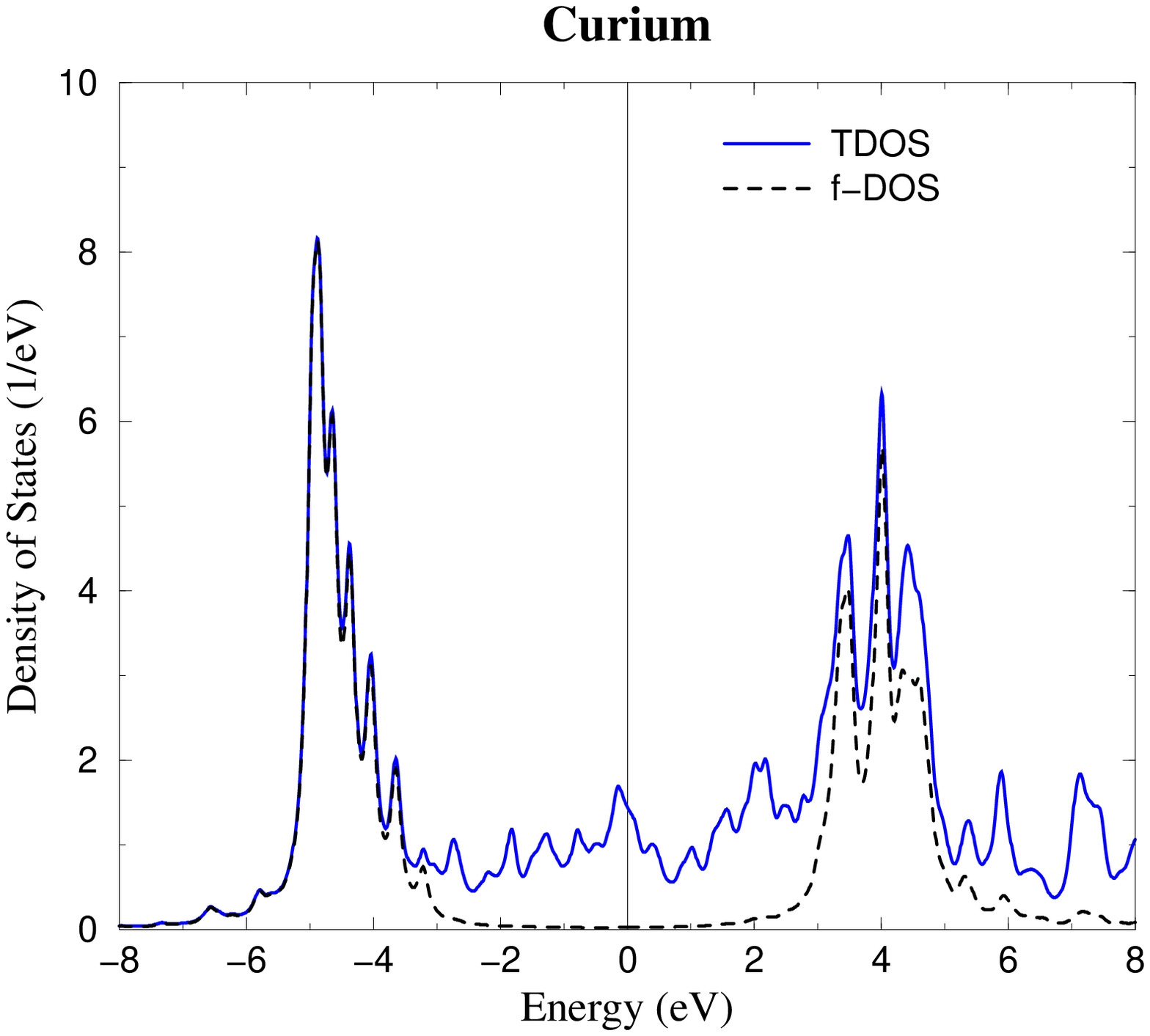}
\includegraphics[width=3.25in,clip]{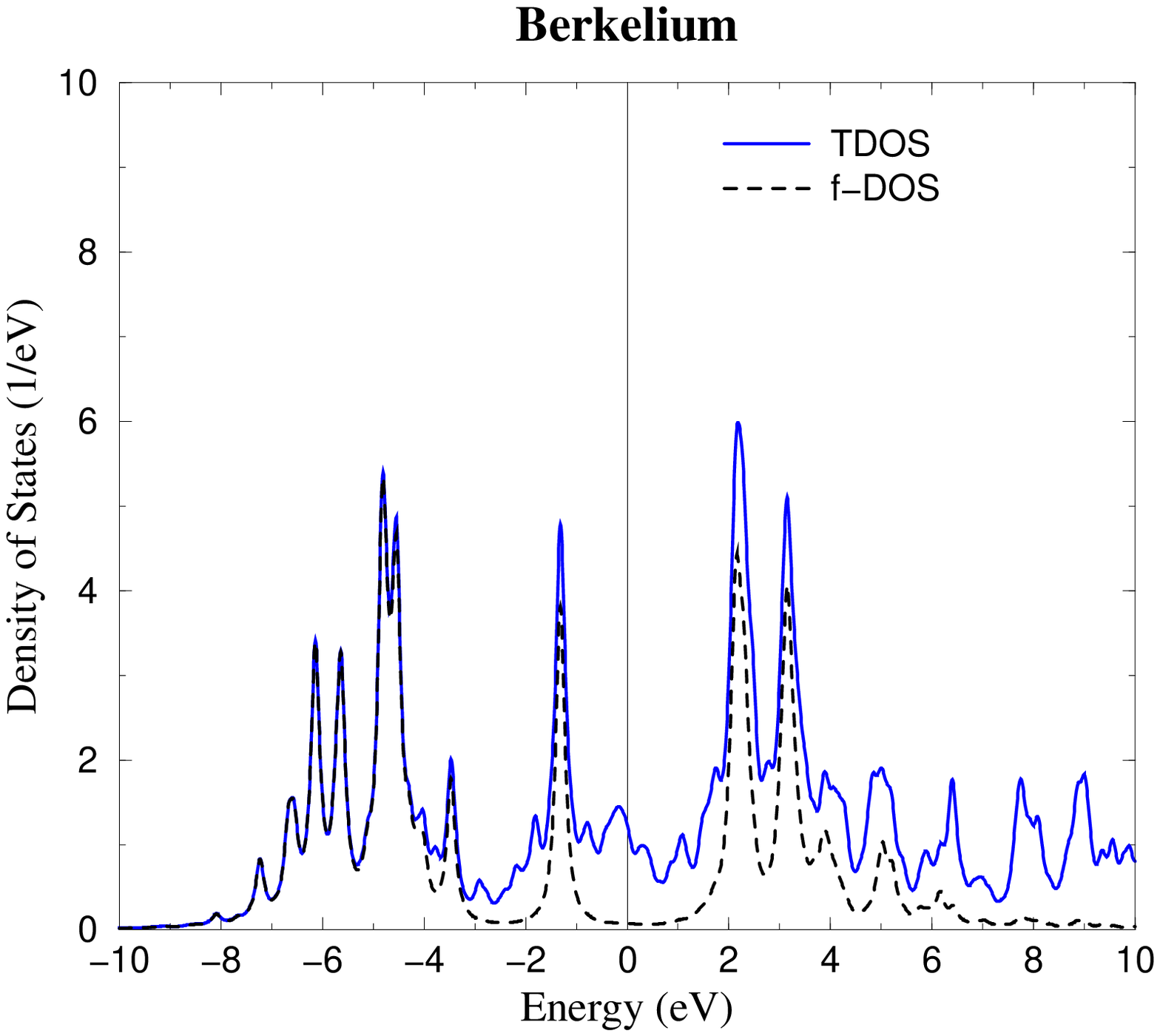}
\caption{\label{fig:spectra}Total-DOS and $f$-DOS for $fcc$-Am,
  $fcc$-Cm, and $fcc$-Bk for $\beta$= 100 eV$^{-1}$.}
\end{figure}

We plot in Fig.~\ref{fig:spectra} total and $f$-projected spectral
densities resulting from self-consistent LDMA calculations (i.~e.,
from converged Eq.~\eqref{eq:16}). In the case of Am, we obtain very
good agreement with previous LDA+DMFT calculations \cite{Svane06,
Savrasov06} as well as with our own non-self-consistent LDA+HIA
calculations \cite{shick07} for spectral peak positions in occupied
and unoccupied parts of the spectrum. Comparison with the PES
experimental data \cite{Naegele} is very good. No PES and BIS
measurements exist for Cm and Bk metals. The results of present Cm
calculations agree reasonably well with the results of recent DMFT
study \cite{gaby} supporting validity of the LDMA. We found
practically no changes in the densities of states when $\beta$ was
decreased from 10 eV$^{-1}$ to 100 eV$^{-1}$. The results turn out
to be almost insensitive to the choice of $\beta$, since it enters
explicitly only Eq.~(\ref{eq:12}) and has practically no influence
on the chemical potential $\mu_H$.

For Am and Cm, the self-consistent value of $n_f$ is very close to
the atomic integer value (see Tab.~\ref{tab:occupations}) in
agreement with LDA+DMFT results \cite{Savrasov06,gaby}. For Bk,
deviation of $n_f$ from the nominal atomic $f^8$ is somewhat bigger
(Tab.~\ref{tab:occupations}), suggesting a possibility of
mixed-valence character in some of heavy actinides. In fact, Svane
{\itshape et al.}~\cite{SIC} have already suggested mixed-valence
states in Am, Cm and Bk on the basis of SIC-LSD calculations (see
Tab.~I of Ref.~\cite{SIC}) that split the $f$-electrons into
localized manifold with fixed valence and an itinerant part. Present
LDMA as well as  previous LDA+DMFT \cite{Savrasov06,gaby}
calculations show that the tendency to mixed-valence in heavy
actinides is substantially overestimated by the SIC-LSD theory.

Now we turn to comparison with XAS and EELS experiments
\cite{Moore07}. In these experiments, the intensities $I_{5/2}$
($4d_{5/2} \rightarrow 5f_{5/2,7/2}$) and $I_{3/2}$ ($4d_{3/2}
\rightarrow 5f_{5/2}$) of the X-ray absorption lines are measured and
the branching ratio $B = I_{5/2}/(I_{3/2} + I_{5/2})$ is obtained.
Note that $B$ is the only quantity which directly follows from the
experiments. To extract the SOC strength $w^{110}$, the
atomic sum rules are used in conjunction with the atomic
calculations \cite{Moore}. In order to compare with the experiment,
we obtain $n_{5/2}$ and $n_{7/2}$ from the local occupation matrix
$n_{\gamma_1 \gamma_2}$ and make use of Eq.~(\ref{eq:BR}) to obtain
$B$ and ${w^{110}}$. We do not take into account the small correction
factor $\Delta$ \cite{Moore}.

The LDMA results for Am are shown in Table~\ref{tab:occupations}
in comparison with the experimental
data \cite{Moore07a} and the  results of atomic
intermediate-coupling (IC) calculations \cite{IC}. The LDMA
calculated $n_{5/2}$, $n_{7/2}$, branching ratio $B$, and spin-orbit
coupling strength are close to atomic IC and experimentally derived
values. Once again, present calculations confirm localized nature of
solid state Am $f$-manifold close to the atomic $f^6$
configuration.

The LDMA results for Cm are also shown in
Table~\ref{tab:occupations} in comparison with the results of DMFT
calculations \cite{gaby}, atomic IC calculations \cite{IC} as well
as with experimental data \cite{Moore07}. There is a very good
agreement for $n_{5/2}$, $n_{7/2}$,  $B$, and ${w^{110}}/{(14 -
n_f)}$ between LDMA and atomic IC calculations. Also, the calculated
branching ratio agrees with $B$=0.75 obtained from DMFT calculations
\cite{gaby}. Note that LDMA, DMFT and IC results agree with each
other, and all slightly differ from experimentally observed $B$  of
0.794 \cite{Moore07}. Gaining inspiration from LDA/GGA, Moore
{\itshape et al.}~\cite{Moore07} suggested that Cm 5$f$-states are
shifted towards the LS coupling limit due to enhancement of the
exchange interaction over the spin-orbit coupling. However, Shim
{\itshape et al.}~\cite{ShimEPL}  noticed very recently that
agreement between the theory and experiment for $B$ improves
substantially when the Slater integrals \cite{Moore} are slightly
reduced to account for the solid state screening.

To date, no XAS or EELS experimental data exist for Bk metal. The
calculated  $n_{5/2}$, $n_{7/2}$, branching ratio $B$, and spin-orbit
coupling strength are listed in Table~\ref{tab:occupations} together
with the atomic IC
$f^8$ calculations.  The  main difference between the solid state
and the atomic $f$-manifolds is due to an increase in occupation of
$n_{7/2}$ states. Nevertheless, the values of $B$ and $w^{110}$ per
hole are practically the same. The measurements of the branching
ratio are often used to obtain the experimental value of $n_f$.
Our results illustrate that the knowledge of the $B$-ratio alone is
not sufficient for precise determination of the $f$-manifold
occupation.

\begin{table}[h]
\begin{ruledtabular}
\begin{tabular}{lcccccc}
{\bf Am}  & $n_f$  & $n_f^{5/2}$ &  $n_f^{7/2}$& $B$  &${w^{110}}/{n_h}$ \\
\hline
LDMA ($\beta$= 10 eV$^{-1}$)  & 5.95 & 5.11 & 0.83 & 0.897 & -0.743 \\
LDMA ($\beta$= 100 eV$^{-1}$)  & 5.95 & 5.16 & 0.79 & 0.902 & -0.756 \\
atomic IC \cite{Moore}& 6 & 5.28 & 0.72 & 0.916 & -0.79 \\
Exp. \cite{Moore07} & 6 & 5.38 & 0.62 & 0.930 & -0.825 \\
\hline
{\bf Cm}  & $n_f$  & $n_f^{5/2}$ &  $n_f^{7/2}$& $B$  & ${w^{110}}/{n_h}$ \\
\hline
LDMA ($\beta$= 10 eV$^{-1}$) & 7.07 & 4.04 & 3.03 & 0.736 & -0.340 \\
LDMA ($\beta$= 100 eV$^{-1}$) & 7.07 & 4.04 & 3.03 & 0.737 & -0.341 \\
DMFT \cite{gaby} & 7.0 & N/A & N/A & 0.75 & N/A \\
atomic IC \cite{Moore}& 7 & 4.10 & 2.90 & 0.75 & -0.37 \\
Exp. \cite{Moore07} & 7 & 4.41 & 2.59 & 0.794 & -0.485 \\
\hline
{\bf Bk} & $n_f$  & $n_f^{5/2}$ &  $n_f^{7/2}$& $B$  & ${w^{110}}/{n_h}$ \\
\hline
LDMA ($\beta$= 10 eV$^{-1}$)& 8.22 & 5.01 & 3.21 & 0.840 & -0.591 \\
LDMA ($\beta$= 100 eV$^{-1}$)& 8.22 & 5.01 & 3.21 & 0.840 & -0.601 \\
atomic IC \cite{Moore}& 8 & 5.00 & 3.00 & 0.84 & -0.61 \\
\end{tabular}
\end{ruledtabular}
\caption{Branching ratio $B$ and spin-orbit coupling strength per
hole ${w^{110}}/n_h$, where $n_h={(14 - n_f)}$, for Am,
Cm and Bk. Note that ``experimental" values of $n_f^{5/2}$ and $n_f^{7/2}$ are
not measured, but derived from sum rule Eq.~(\ref{eq:BR}) assuming
integer atomic occupation $n_f$.} \label{tab:occupations}
\end{table}

Now we turn to an estimate of the effective local magnetic moment
$\mu_{eff}$ in the paramagnetic phase. Importance of magnetism in Cm
metal was emphasized recently in the context of its phase stability
\cite{Heathman05}. The temperature independent magnetic
susceptibility is found for Am \cite{Hurray} that is consistent with
zero $\mu_{eff}$. The magnetic susceptibility measurements in
the paramagnetic phase yield effective magnetic moment of $\sim$ 8
$\mu_B$ for Cm and $\sim$ 9.8 $\mu_B$ for Bk \cite{Hurray}.

We can estimate semi-quantitatively the effective local moment
making use of the atomic Hamiltonian, Eq.~(\ref{eq:10}), and the
chemical potential $\mu_H=V_{dc}$ that is self-consistently
determined in the LDMA calculations. The expectation values of
total, spin and orbital moment operators, ${\mathbf J}$, ${\mathbf
S}$ and ${\mathbf L}$, are calculated as grand-canonical averages,
\begin{eqnarray}
\langle {\mathbf X}^2 \rangle &=& \frac{1}{Z} {\rm Tr} \Big[
{\mathbf X}^2 \exp(-\beta [H^{\rm at}
       - \mu_H \hat{N}]) \Big] \; \, , \nonumber \\
{\mathbf X} &=& {\mathbf J, \mathbf S, \mathbf L} \; .
\label{eq:mmom}
\end{eqnarray}
Further, spin $S$, orbital $L$ and total $J$ moment ``quantum
numbers'' are found using $\langle {\mathbf X}^2 \rangle = X(X+1)$
for $ X=S,L,J$. Subsequently, the effective magnetic moment
$\mu_{eff} = g_J \sqrt{J(J+1)}$ is evaluated, where the g-factor
$g_J = (2S+L)/J$ is used.

For Am, we obtain $S=L=2.33$ and $J=0$ for $\beta = 100$ eV$^{-1}$
in Eq.~(\ref{eq:mmom}). Decreasing the value of $\beta$ to
10~eV$^{-1}$ yields a small difference in $S$ and $L$ values, and
gives a non-zero value of $J= 0.099$. It means that the thermal
population of the multiplets excited over the non-magnetic $f^6$
ground state starts to produce non-negligible contribution in
Eq.~(\ref{eq:mmom}).

For Cm, $S=3.30$, $L=0.40$, and $J=3.50$  are calculated from
Eq.~(\ref{eq:mmom}) for $\beta = 100$~eV$^{-1}$. Decrease of $\beta$
to 10~eV$^{-1}$ produces practically no difference in $S$, $L$, and
$J$ values.
The corresponding local magnetic moment $\mu_{eff}= 7.94$ $\mu_B$
agrees well with atomic IC value and experimental data \cite{Hurray}
(see Tab.~\ref{tab:spins}).

For Bk, Eq.~(\ref{eq:mmom}) yields $S=2.71$, $L=0.40$, and
$J=6.00$ for $\beta = 100$~eV$^{-1}$, as well as for $\beta =
10$~eV$^{-1}$. The effective magnetic moment $\mu_{eff}= 9.8$ $\mu_B$
agrees well with the atomic $f^8$ IC value and experimental data
\cite{Hurray} shown in Table~\ref{tab:spins}.

Our calculations, which are not bound by any particular atomic
coupling scheme, illustrate once again that IC scheme is suitable
for heavy actinides. Also, a good agreement of estimated $\mu_{eff}$
with experimental data is somewhat surprising.

\begin{table}[h]
\begin{ruledtabular}
\begin{tabular}{lcccccc}
  $\mu_{eff}$($\mu_B$)   & Am  & Cm   &  Bk    \\
\hline
  LDMA         & 0   & 7.94 & 9.54    \\
  IC \cite{Hurray}  & 0   & 7.6 &  9.3     \\
Exp.\cite{Hurray}   & 0   & $\sim$ 8 & $\sim$ 9.8  \\
\end{tabular}
\end{ruledtabular}
\caption{Effective local magnetic moment $\mu_{eff}$ for Am, Cm and
Bk. The atomic IC vales of $\mu_{eff}$ and experimental data
\cite{Hurray} are shown.} \label{tab:spins}
\end{table}

\section{Discussion and Conclusions}
For a better insight, it is useful to point out that in the current
implementation, which is based on a single-site approximation,
Eq.~(\ref{eq:16}), to the solution of Eq.~(\ref{eq:1gf}),
the LDMA can be regarded as an extension of the
LDA+U. Importantly, the on-site occupation matrix $n_{\gamma_1
\gamma_2}$ is now evaluated in a many-body Hilbert space rather than
in a single-particle Hilbert space as in the conventional LDA+U
\cite{LAZ95}. Current implementation can be further extended towards
a fully self-consistent DMFT making use of Wannier-like basis set
together with more sophisticated approximation for the quantum
impurity solver along the lines proposed in Ref.~\cite{Amadon}.

Our approach to the charge self-consistency is essentially different
from the one proposed by Lechermann {\itshape et al.} \cite{Lechermann}.  The
on-site occupation matrix, instead of the full charge density, is obtained
from the local Green function. The corresponding orbital-dependent
effective potential is used in Eq.~(\ref{eq:kseq}) to calculate a new
bath Green function ${G}_{LDA}$, Eq.~(\ref{eq:15}), instead of
orbital-independent Kohn-Sham (LDA/GGA) potential.

In this paper, we do not address the very important issues of the total
energy calculation and determination of the equilibrium lattice
properties. The practical implementation of accurate total energy
calculations is ongoing work that will be discussed in detail in the
future.

To summarize, we have presented a straightforward and numerically
efficient local density matrix approximation (LDMA) to perform the LDA+HIA
calculations in the FP-LAPW basis, including self-consistency over
the charge density. This implementation is all-electron, incorporates
spin-orbit interaction, and includes no shape approximations for the
charge density. The method works well for the electronic spectrum of
representative actinide Am, Cm and Bk metals. Importantly, the method
allows fully self-consistent calculations for the paramagnetic phase
of the local moment systems with strong Coulomb correlations.
It can be extended to incorporate the total energy and to treat the
magnetically ordered phases.

We are grateful to V. Drchal and V. Jani\u{s} for helpful comments
and discussion. This work was supported by the Grant Agency of Czech
Republic (Project 202/07/0644) and German-Czech collaboration
program (436TSE113/53/0-1, GACR 202/07/J047).

\begin{thebibliography}{99}
\bibitem{SK} S. Y. Savrasov, and G. Kotliar, Phys. Rev. Lett. {\bf 84}, 3670
(2000).
\bibitem{shick05} A. B. Shick, V. Drchal, and L. Havela, Europhys.
                  Lett. {\bf 69}, 588 (2005).
\bibitem{HYF} D. Torumba, P. Novak, and S. Cottenier, Phys. Rev. B {\bf 77}, 155101 (2008).
\bibitem{SIC} A. Svane, L. Petit, Z. Szotek, and W. M. Temmerman,
  Phys. Rev. B {\bf 76}, 115116 (2007).
\bibitem{gaby01} S.Y. Savrasov, G. Kotliar, and E. Abrahams, Nature {\bf
410}, 793 (2001).
\bibitem{gaby}J. H. Shim, K. Haule, and G. Kotliar, Nature {\bf
446}, 513 (2007).
\bibitem{shick07} A. Shick, J. Koloren\v{c}, L.
Havela, V. Drchal, and T. Gouder, Europhys. Lett. {\bf 17}, 17003 (2007).
\bibitem{Zhu} J.-X. Zhu, A. K. McMahan, M. D. Jones, T. Durakiewicz,
  J. J. Joyce, J. M. Wills, and R. C. Albers, Phys. Rev. B {\bf 76},
245118 (2007).
\bibitem{Cris} C. A. Marianetti, K. Haule, G.
Kotliar, and M. J. Fluss, Phys. Rev. Lett. {\bf 101}, 056403 (2008).
\bibitem{Wimmer} E. Wimmer, H. Krakauer, M. Weinert, and A. J. Freeman,
Phys. Rev. B {\bf 24}, 864 (1981).
\bibitem{Moore} K. T. Moore and G. van der Laan, Rev. Mod. Phys. {\bf 81}, 235 (2009).
\bibitem{Griveau} J.-C. Griveau, J. Rebizant, G. H. Lander, and G.
Kotliar, Phys. Rev. Lett. {\bf 94}, 097002 (2005).
\bibitem{Heathman05} S. Heathman, R. G. Haire, T. Le Bihan,
  A. Lindbaum, M. Idiri, P. Normile, S. Li, R. Ahuja, B. Johansson,
  and G. H. Lander, Science {\bf 309}, 110 (2005).
\bibitem{Moore07} K. T. Moore, G. van der Laan, R. G. Haire,
  M. A. Wall, A. J. Schwartz, and P. S\"{o}derlind,
  Phys. Rev. Lett. {\bf 98}, 236402 (2007).
\bibitem{LK99} A. I. Lichtenstein and M. I. Katsnelson, Phys. Rev. B {\bf
57}, 6884 (1998).
\bibitem{LAZ95} A. I. Liechtenstein, V. I. Anisimov,  and J. Zaanen,
Phys. Rev. B {\bf 52}, R5467 (1995).
\bibitem{shift} H. Kajueter and G. Kotliar, Phys. Rev. Lett. {\bf 77}, 131 (1996).
\bibitem{shick2004}
A. B. Shick, V. Jani\v{s}, V. Drchal, and W. E. Pickett, Phys. Rev.
B {\bf 70}, 134506 (2004).
\bibitem{SLP99} A. B. Shick, A. I. Liechtenstein, and W. E. Pickett,
Phys. Rev. B {\bf 60}, 10763 (1999).
\bibitem{Solovyev} I. V. Solovyev,  P. H. Dederichs, and V. I. Anisimov,
Phys. Rev. B {\bf 50}, 16861 (1994).
\bibitem{Anisimov} V. I. Anisimov, J. Zaanen, and O. K. Andersen, Phys.
Rev. B {\bf 44}, 943 (1991).
\bibitem{Kunes} J. Kune\v{s}, V. I. Anisimov, A. V. Lukoyanov, and D.
Vollhardt, Phys. Rev. B {\bf 75}, 165115 (2007).
\bibitem{Georges} A. Georges, G. Kotliar, W. Krauth, and M.
Rozenberg, Rev. Mod. Phys. {\bf 68}, 13 (1996).
\bibitem{Savrasov06} S. Y. Savrasov, K. Haule, and G. Kotliar, Phys. Rev. Lett. {\bf 96},
036404 (2006).
\bibitem{BZ}
H. J. Monkhorst and J. D. Pack, Phys. Rev. B. {\bf 13}, 5188 (1976).
\bibitem{Svane06} A. Svane, Solid State Commun. {\bf 140}, 364 (2006).
\bibitem{Naegele} J. R. Naegele, L. Manes, J. C. Spirlet, and
  W. M\"uller, Phys. Rev. Lett. {\bf 52}, 1834 (1984).
\bibitem{Moore07a} K. T. Moore, G. van der Laan, M. A. Wall,
  A. J. Schwartz, and R. G. Haire, Phys. Rev. B {\bf
76}, 073105 (2007).
\bibitem{IC} Here we took the IC $n_{5/2}$, $n_{7/2}$ values given in Table
IV of Ref.~\cite{Moore} and used Eq.~(\ref{eq:BR}) to calculate
$w^{110}$ and $B$.
\bibitem{ShimEPL} J. H. Shim, K. Haule, and G. Kotliar,
Europhys. Lett. {\bf 85}, 17007 (2009).
\bibitem{Hurray} P. G. Huray and S. E. Nave, in
Handbook on the Physics and Chemistry of the Actindes, Vol. 5,
Freeman, A. J. and Lander, G. H., Eds. (Elsevier, Amsterdam) p. 311
(1987).
\bibitem{Amadon}  B. Amadon, F. Lechermann, A. Georges, F. Jollet,
  T. O. Wehling, and A. I.
Lichtenstein, Phys. Rev. B {\bf 77}, 205112 (2008).
\bibitem{Lechermann}  F. Lechermann, A. Georges, A. Poteryaev,
  S. Biermann, M. Posternak, A. Yamasaki, and O. K.
Andersen, Phys. Rev.  B {\bf 74}, 125120 (2006).
\end{thebibliography}
\end{document}